\documentclass[12pt,fleqn]{article}
\usepackage{epsfig}
\usepackage[dvips]{color}
\begin{document}
\title{Electroweak phase transition in MSSM with $U(1)'$ in explicit CP violation scenario}
\author{S.W. Ham$^{(1)}$ and S.K. Oh$^{(1,2)}$
\\
\\
{\it $^{(1)}$ Center for High Energy Physics, Kyungpook National University}
\\
{\it Daegu 702-701, Korea}
\\
{\it $^{(2)}$ Department of Physics, Konkuk University, Seoul 143-701, Korea}
\\
\\
}
\date{}
\maketitle
\begin{abstract}
The possibility of a strongly first-order electroweak phase transition is established
in the minimal supersymmetric standard model with an extra $U(1)'$,
where a nontrivial CP violating phase is introduced in its Higgs sector.
We find that there is a wide region in the parameter space of the model that allows the
strongly first-order electroweak phase transition.
The mass of stop quark need not be smaller than the top quark mass
to ensure the first-order electroweak phase transition be strong.
The effect of the CP violating phase upon the strength of the phase transition is discovered.
The strength of the phase transition is reduced when the size of the CP violation is increased.
For a given CP violating phase, we find that the model has a larger mass for the
lightest Higgs boson when it has a stronger phase transition.
\end{abstract}

\vfil\eject

\section{INTRODUCTION}

The three necessary conditions for explaining the observed baryon asymmetry of the universe,
suggested several decades ago by Sakharov [1], are the violation of baryon number conservation,
the violation of both C and CP, and the deviation from thermal equilibrium.
As is well known, in order to ensure sufficient deviation from thermal equilibrium,
the electroweak phase transition (EWPT) that is responsible for the baryogenesis should be strongly first-order,
since otherwise the baryon asymmetry generated during the EWPT would subsequently disappear.
The criterion for strength of the EWPT is given
by the baryon preservation condition $v_c(T_c) \ge T_c$ [2], where $v_c(T_c)$ and $T_c$ are
respectively the critical vacuum expectation value (VEV) and the critical temperature.

In the literature, a large number of articles have been devoted to the possibility of establishing the strongly first-order EWPT in various theoretical models, including the Standard Model (SM) which is naturally the primary candidate.
It has been found however that the SM is inadequate to accommodate the desired EWPT, because it cannot produce the CP violation large enough to generate the baryon asymmetry by means of the complex phase in the Cabibbo-Kobayashi-Maskawa matrix alone [3,4], and it cannot make the EWPT strong enough unless the mass of the SM Higgs boson is below the present experimental lower bound.

Attentions have been paid to alternative models beyond the SM within the context of the electroweak baryogenesis.
Those models embedded in the supersymmetry (SUSY), having at least two Higgs doublets, are considered as the most plausible alternatives since they can bring about additional sources of CP violation in their Higgs sectors [5].
Among them, it has been shown that the minimal  supersymmetric standard model (MSSM) with a light stop quark, being consistent with the present experimental bounds for both the Higgs bosons and the supersymmetric particles, allows the possibility of electroweak baryogenesis [6-14].
The tree-level Higgs potential of the non-minimal supersymmetric standard models, extended from the MSSM by introducing an additional Higgs singlet, may possess trilinear terms in Higgs fields.
It has been studied that these trilinear terms may lead the models to the electroweak baryogenesis satisfying the baryon preservation condition in a wide parameter space, without requiring a light stop quark [15-21].

We have also studied several models for the possibility of the strongly first-order EWPT [22-24].
Recently, an MSSM with $U(1)'$ gauge symmetry has been our subject [24].
This model is found to satisfy the baryon preservation for a wide parameter space, without requiring scalar-pseudoscalar mixing or a light stop quark below top quark mass.
This model may also accommodate new sources of CP violation in its SUSY soft terms, which allow enough baryon creation.
It is mainly because its tree-level Higgs potential possesses the trilinear terms like the non-minimal supersymmetric models.

In this paper, we continue to study the MSSM with $U(1)'$, this time including the explicit CP violation.
In the explicit CP violation scenario, we examine carefully whether the strongly first-order EWPT may be realized in the model.
We take into account the radiative corrections due to top and stop quarks, with the scalar-pseudoscalar mixings.
We find that a strongly first-order EWPT is indeed viable for some parameter space of this model in explicit CP violation scenario.
Then, we investigate the effects of the CP mixing between scalar and pseudoscalar Higgs bosons, which is produced by the CP phase in the stop quark masses, on the strength of the first-order EWPT.
We find that the CP phase arising from the stop quark masses reduces slightly the strength of the first-order EWPT.
On the other hand, the lightest Higgs boson mass does not seriously affect the strength of the first-order EWPT, since it may be either enhanced or reduced by increasing the lightest Higgs boson mass, depending on other parameters.

\section{HIGGS SECTOR AT ZERO TEMPERATURE}

The Higgs sector of the MSSM with $U(1)'$ consists of two Higgs doublets $H_1 = (H_1^0, H^-)$ and $H_2 = (H^+, H_2^0)$, and a neutral Higgs singlet $S$ [25-34].
Retaining only the Yukawa coupling of top quark, the superpotential of this model may be written as
\begin{equation}
    W \approx h_t Q H_2 t_R^c + \lambda N H_1^T \epsilon H_2 \  ,
\end{equation}
where $h_t$ is the top Yukawa coupling coefficient, $Q$ is the left-handed quark doublet superfield of the third generation, $t_R^c$ is the charge conjugate of the right-handed top quark superfield, $\lambda$ is the dimensionless coupling constant, and $\epsilon$ is a $2 \times 2$ antisymmetric matrix defined as $\epsilon_{12} = 1$.

The tree-level Higgs potential of this model at zero temperature may be decomposed into $F$-terms, $D$-terms, and soft terms as
\begin{equation}
V_0 = V_F + V_D + V_{\rm S} \ ,
\end{equation}
with
\begin{eqnarray}
V_F & = & |\lambda|^2 [(|H_1|^2 + |H_2|^2) |S|^2 + |H_1^T \epsilon H_2|^2]  \ , \cr
V_D & = & {g_2^2 \over 8} (H_1^{\dagger} \vec\sigma H_1 + H_2^{\dagger} \vec\sigma H_2)^2
+ {g_1^2 \over 8} (|H_1|^2 - |H_2|^2)^2 \cr
& &\mbox{}+ {g^{'2}_1 \over 2} ( {\tilde Q}_1 |H_1|^2 + {\tilde Q}_2 |H_2|^2 + {\tilde Q}_3 |S|^2)^2 \ , \cr
V_{\rm S} & = & m_1^2 |H_1|^2 + m_2^2 |H_2|^2 + m_3^2 |S|^2
- [\lambda A_{\lambda} H_1^T \epsilon H_2 S + {\rm H.c.}] \ ,
\end{eqnarray}
where $g_2$, $g_1$, and $g'_1$ are respectively the gauge coupling coefficients of $SU(2)$, $U(1)$, and $U(1)'$,
$\vec\sigma$ are Pauli matrices, $A_{\lambda}$ is the trilinear soft SUSY breaking parameter with mass dimension, $m_i$ ($i$ = 1-3) are soft SUSY breaking masses, and ${\tilde Q}_i$ ($i$ = 1-3) are respectively the effective $U(1)'$ charges of $H_1$, $H_2$, and $S$.

The effective $U(1)'$ charges of the Higgs fields satisfy the identity relation of $\sum_{i=1}^3 {\tilde Q}_i = 0$ in order to ensure the $U(1)'$ gauge invariance.
The soft SUSY breaking masses are then eliminated by the three minimum conditions of the Higgs potential.
The three minimum conditions define the vacuum and the vacuum expectation values
of the three neutral Higgs fields at zero temperature, namely,
$v_1 = \langle H_1^0 \rangle$, $v_2 = \langle H_2^0 \rangle$,
and $s e^{i \phi_s}= \langle S \rangle$, where we assume that the Higgs singlet $S$ may develop
a complex VEV.

Besides the phase $\phi_s$, the Higgs potential may explicitly contain complex coefficients.
While other terms can be made real through redefinition of the Higgs fields, the trilinear terms, $\lambda A_{\lambda} H_1^T \epsilon H_2 S +{\rm H.c.}$, may remain complex.
Thus, we are left with a complex phase $\phi_0 = \phi_{\lambda} + \phi_{A_{\lambda}} + \phi_s$, where $\phi_{\lambda}$ is the phase of $\lambda$ and  $\phi_{A_{\lambda}}$ is that of $A_{\lambda}$.
At the tree level, this complex phase $\phi_0$ can always be eliminated by rotating suitably the relevant Higgs fields.
The Higgs potential does not yield any CP mixing among the scalar and pseudoscalar Higgs fields [33,34].

At the one-loop level, a non-trivial complex phase emerges in the stop quark masses, which yields the mixing of the scalar and pseudoscalar Higgs bosons.
In general, the contributions from the top and stop quark loops are most dominant at the one-loop level for a wide parameter space.
The one-loop effective potential, at zero temperature, due to the top and stop quark loops is given as [35]
\begin{equation}
V_1(0) = \sum_{i = 1}^2 {n_{{\tilde t}_i} {\cal M}_{{\tilde t}_i}^4 \over 64 \pi^2}
  \left (\log {{\cal M}_{{\tilde t}_i}^2 \over \Lambda^2} - {3\over 2} \right )
  + {n_t {\cal M}_t^4 \over 64 \pi^2} \left (\log {{\cal M}_t^2 \over \Lambda^2}
  - {3\over 2} \right ) \ ,
\end{equation}
where $\Lambda$ is the renormalization scale in the modified minimal subtraction scheme, and $n_t = -12$ and $n_{{\tilde t}_i} = 6$ ($i=1,2$) are respectively the degrees of freedom for top and stop quarks from color, charge, and spin factors.

We assume that the left-handed and right-handed stop quarks are not degenerate.
Then, the stop quark masses are given as
\begin{equation}
m_{{\tilde t}_1, {\tilde t}_2}^2 = m_T^2 + m_t^2 \mp
\sqrt{h_t^2 A_t^2 v_2^2 + h_t^2 \lambda^2 v_1^2 s^2 - 2 h_t^2 \lambda A_t v_1 v_2 s \cos \phi_t}
\end{equation}
where $m_T$ is the soft SUSY breaking mass, $m_t = h_t v_2$ is the mass of top quark, $A_t$ is the trilinear SUSY breaking parameter with mass dimension, and $\phi_t$ is a phase originating from the phases of $\lambda$, $A_t$, and $s$.

In case of explicit CP violation, at the one-loop level, the non-trivial tadpole minimum condition with respect to the pseudoscalar component of the Higgs field is given as
\begin{equation}
0 = A_{\lambda} \sin \phi_0
- {3 m_t^2 A_t \sin \phi_t \over 16 \pi^2 v^2 \sin^2 \beta} f (m_{{\tilde t}_1}^2,  \ m_{{\tilde t}_2}^2) \ ,
\end{equation}
where $\phi_0$, which can be eliminated by rotating the neutral Higgs fields at the tree level, can not be so at the one-loop level, and the dimensionless function $f$ is defined as
\[
 f(m_x^2, \ m_y^2) = {1 \over (m_y^2 - m_x^2)} \left[  m_x^2 \log {m_x^2 \over \Lambda^2} - m_y^2
\log {m_y^2 \over \Lambda^2} \right] + 1 \ .
\]

In the present model, there are four neutral Higgs fields.
The physical Higgs bosons and their squared masses are given respectively by the eigenvectors and the eigenvalues, denoted as $m^2_{h_i}$ ($i$ = 1-4), of the $4 \times 4$ mass matrix for the neutral Higgs fields.
We sort these four neutral Higgs bosons in the increasing order of their masses such that $m^2_{h_1}$ is the smallest eigenvalue and $h_1$ is the lightest neutral Higgs boson.
The explicit expression for the mass matrix may be found elsewhere [33,34].

The elements of the mass matrix, which are responsible for the mixing between the scalar and pseudoscalar Higgs bosons, are proportional to $\sin \phi_t$.
Thus, the CP mixing in the present model at the one-loop level is triggered by $\phi_t$, which is practically the only source of CP violation in this model.
If $\phi_t = 0$, these four neutral Higgs bosons may be classified into three scalar and a pseudoscalar Higgs bosons.

\section{THERMAL EFFECTS IN HIGGS POTENTIAL}

Up to now, we have studied the Higgs potential of the MSSM with $U(1)'$ at zero temperature.
We now include the thermal effects to the Higgs potential in order to examine its behavior at the finite temperature.
For the thermal effects, we take into account the one-loop contributions due to the top and stop quarks at finite temperature $T$, which is given as [36]
\begin{eqnarray}
V_1(T) & = & {n_t T^4 \over 2 \pi^2}
            \int_0^{\infty} dx \ x^2 \
            \log \left [1 + \exp{\left ( - \sqrt {x^2+{{\cal M}_t^2/T^2 }} \right )  } \right ] \cr
& & + \sum_{i = 1, 2} {n_{{\tilde t}_i} T^4 \over 2 \pi^2}
            \int_0^{\infty} dx \ x^2 \
            \log \left [1 - \exp{\left ( - \sqrt {x^2+{{\cal M}_{{\tilde t}_i}^2/T^2 }} \right )  } \right ] \ ,
\end{eqnarray}
where the first term represents the thermal effects of top quark, and second term those of the stop quarks, and the temperature dependence is explicitly shown.
The masses of the field-dependent stop quarks in the above expression contain implicitly $\phi_t$ that triggers the CP violation.
The full effective potential at the one-loop level at finite temperature is therefore $V(T) = V_0 + V_1 (0) + V_1 (T)$.

In order to study the possibility of a strongly first-order EWPT in this model, we examine if the full effective potential may develop two degenerate vacua at finite temperature, which is the critical temperature for the phase transition, and also examine if it may satisfy the baryon preservation condition.
It is known that the trilinear terms proportional to $A_{\lambda}$ in the tree-level Higgs potential of the present model allows a strongly first-order EWPT for a wide parameter space, if the CP violation is turned off in its Higgs sector [34].
Then, we are mainly interested in the behavior of the present model in the presence of an explicit CP violation in its Higgs sector.
In particular, we study the dependence of the strength of the strongly first-order EWPT upon the size of CP violation.

In terms of the VEVs at finite temperature, the vacuum is determined by the minimum of the following effective potential,
\begin{equation}
\langle V(v_1,v_2,s,T) \rangle = \langle V_0 \rangle + \langle V_1 (0) \rangle + \langle V_1 (T) \rangle \ ,
\end{equation}
where $v_1$, $v_2$, and $s$ are now the temperature-dependent vacuum expectation values with
\begin{eqnarray}
\langle V_0 \rangle & = & m_1^2 v_1^2 + m_2^2 v_2^2 + m_3^2 s^2 + {g_1^2 + g_2^2 \over 8} (v_1^2 - v_2^2)^2
+ \lambda^2 (v_1^2 v_2^2 + v_1^2 s^2 + v_2^2 s^2)  \cr
& &\mbox{} - 2 \lambda A_{\lambda} v_1 v_2 s \cos \phi_0 + {g^{'2}_1 \over 2 }
({\tilde Q}_1 v_1^2 + {\tilde Q}_2 v_2^2 + {\tilde Q}_3 s^2)^2 \ , \cr
\langle V_1 (0) \rangle & = & \sum_{i = 1}^2 {3 m_{{\tilde t}_i}^4 \over 32 \pi^2}
  \left (\log {m_{{\tilde t}_i}^2 \over \Lambda^2} - {3\over 2} \right )
  - {3 m_t^4 \over 16 \pi^2} \left (\log { m_t^2 \over \Lambda^2}
  - {3\over 2} \right ) \ , \cr
\langle V_1 (T) \rangle & = & \mbox{} - {6 T^4 \over \pi^2}
   \int_0^{\infty} dx \ x^2 \
   \log \left [1 + \exp{\left ( - \sqrt {x^2+{m_t^2(v_2)/T^2 }} \right )  } \right ] \cr
& &\mbox{} + \sum_{i=1}^2 {3 T^4 \over \pi^2}
   \int_0^{\infty} dx \ x^2 \
   \log \left [1 - \exp{\left ( - \sqrt {x^2+ m_{{\tilde t}_i}^2(v_1, v_2, s)/T^2 } \right )  } \right ]  \ .
\end{eqnarray}

The minimum conditions for the one-loop effective potential at the zero temperature may express
the soft SUSY masses $m_i$ ($i=1,2,3$) in $\langle V_0 \rangle$ as
\begin{eqnarray}
m_1^2 & = &\mbox{} - {m_Z^2 \over 2} \cos 2 \beta - \lambda^2 (s(0)^2 + v(0)^2 \sin^2 \beta)
                  + \lambda A_{\lambda} s(0) \tan \beta \cos \phi_0  \cr
& &\mbox{} - g^{'2}_1 {\tilde Q}_1 ({\tilde Q}_1 v(0)^2 \cos^2 \beta
+ {\tilde Q}_2 v(0)^2 \sin^2 \beta + {\tilde Q}_3 s(0)^2) \cr
& &\mbox{} + {3 h_t^2 \over 16 \pi^2} (\lambda^2 s(0)^2 - \lambda A_t s(0) \cot \beta \cos \phi_t)
 f(m_{{\tilde t}_1}^2, m_{{\tilde t}_2}^2)   \ , \cr
m_2^2 & = & {m_Z^2 \over 2} \cos 2 \beta - \lambda^2 (s(0)^2 + v(0)^2 \cos^2 \beta)
                  + \lambda A_{\lambda} s(0) \cot \beta \cos \phi_0  \cr
& &\mbox{} - g^{'2}_1 {\tilde Q}_2 ({\tilde Q}_1 v(0)^2 \cos^2 \beta
+ {\tilde Q}_2 v(0)^2 \sin^2 \beta + {\tilde Q}_3 s(0)^2)  \cr
& &\mbox{} + {3 h_t^2 \over 16 \pi^2} (A_t^2 - \lambda A_t s(0) \tan \beta \cos \phi_t)
 f(m_{{\tilde t}_1}^2, m_{{\tilde t}_2}^2)
- { 3 h_t^2 m_T^2 \over 16 \pi^2}
\log \left ( {m_{{\tilde t}_1}^2 m_{{\tilde t}_2}^2 \over \Lambda^4 e^2 } \right ) \cr
& &\mbox{} - { 3 h_t^2 m_t^2 \over 16 \pi^2}
\log \left ( {m_{{\tilde t}_1}^2 m_{{\tilde t}_2}^2 \over m_t^4} \right )
- { 3 h_t^2 \over 32 \pi^2} (m_{{\tilde t}_2}^2 - m_{{\tilde t}_1}^2)
\log \left ( {m_{{\tilde t}_2}^2 \over m_{{\tilde t}_1}^2} \right )  \ , \cr
m_3^2 & = &\mbox{} - \lambda^2 v(0)^2 + {\lambda \over 2 s(0)} v(0)^2 A_{\lambda} \sin 2 \beta \cos \phi_0 \cr
& &\mbox{} - g^{'2}_1 {\tilde Q}_3 ({\tilde Q}_1 v(0)^2 \cos^2 \beta + {\tilde Q}_2 v(0)^2 \sin^2 \beta
+ {\tilde Q}_3 s(0)^2)   \cr
& &\mbox{} + {3 h_t^2 \lambda v(0)^2 \cot \beta \over 16 \pi^2 s(0)}
(\lambda s(0) \cos \beta - A_t \sin \beta \cos \phi_t)
 f(m_{{\tilde t}_1}^2, m_{{\tilde t}_2}^2)  \ ,
\end{eqnarray}
where $v_1 (0)$, $v_2 (0)$, and $s (0)$ are the zero-temperature vacuum expectation values
of the three neutral Higgs fields.
Notice the presence of the phase $\phi_0$, which is expressible as the other parameters
via the tadpole minimum condition at the one-loop level shown in the previous section.

The procedure of exploring the parameter space of the present model starts with reducing the number of variables.
By calculating $\partial V/\partial s|_{v_1, v_2, s} = 0$, which is one of the minimum conditions
that the first derivatives of the effective potential should satisfy, we obtain the following nonlinear equation,
\begin{eqnarray}
0 & = & 2 m_3^2 s - 2 \lambda A_{\lambda} v_1 v_2 \cos \phi_0 + 2 \lambda^2 (v_1^2 + v_2^2) s
  + 2 g^{'2}_1 {\tilde Q}_3 s ({\tilde Q}_1 v_1^2 + {\tilde Q}_2 v_2^2 + {\tilde Q}_3 s^2)  \cr
& & \mbox{} - {3 h_t^2 \lambda v_1 \over 8 \pi^2} (\lambda v_1 s - A_t v_2 \cos \phi_t)
 f(m_{{\tilde t}_1}^2, m_{{\tilde t}_2}^2) \cr
& & \mbox{} - {3 T^2 \over 2 \pi^2} {2 h_t^2 \lambda v_1 \over (m_{{\tilde t}_2}^2 - m_{{\tilde t}_1}^2) }
(\lambda s v_1 - A_t v_2 \cos \phi_t ) \cr
& &\mbox{} \times   \int_0^{\infty} dx \ x^2 \
    {\exp (-\sqrt{x^2 + m_{{\tilde t}_1}^2/T^2 })  \over \sqrt{x^2 + m_{{\tilde t}_1}^2/T^2 }
\{1 - \exp (-\sqrt{x^2 + m_{{\tilde t}_1}^2/T^2 }) \}}  \cr
& &\mbox{} + {3 T^2 \over 2 \pi^2} {2 h_t^2 \lambda v_1 \over (m_{{\tilde t}_2}^2 - m_{{\tilde t}_1}^2) }
(\lambda s v_1 - A_t v_2 \cos \phi_t ) \cr
   & &\mbox{} \times \int_0^{\infty} dx \ x^2
{ \exp (-\sqrt{x^2 + m_{{\tilde t}_2}^2/T^2 })
\over \sqrt{x^2 + m_{{\tilde t}_2}^2/T^2 }
\{1 - \exp (-\sqrt{x^2 + m_{{\tilde t}_2}^2/T^2 }) \}}   \ ,
\end{eqnarray}
which can be used to express $s$ in terms of other variables, $v_1$ and $v_2$.
Thus, by substituting $s$ into $\langle V(v_1,v_2,s,T) \rangle$, we are left with $\langle V(v_1,v_2,T) \rangle$.

For given set of relevant parameter values, we search the minima of $\langle V(v_1,v_2,T) \rangle$ in the $(v_1,v_2)$-plane, varying the temperature.
If two independent but degenerate minima are found for a given temperature, it is the critical temperature, $T_c$, and the shape of the potential at the temperature allows a first-order EWPT.
Let one minimum point in the $(v_1,v_2)$-plane be A, and the other, B.
The distance between A and B is given as
\begin{equation}
v_c = \sqrt{(v_{1B}-v_{1A})^2 +(v_{2B}-v_{2A})^2 + (s_B-s_A)^2} \ ,
\end{equation}
where $s_A$ and $s_B$ are obtained by inverting the above nonlinear equation, by means of the bisection method.
The critical vacuum expectation value, $v_c$, is defined as the distance between the two degenerate minima.
Then, by comparing $v_c$ with the critical temperature, we can check if the baryon preserving condition is satisfied.
In this way, the possibility of a strongly first-order EWPT is established in the MSSM with $U(1)'$, with explicit CP violation.

\section{NUMERICAL ANALYSIS}

Now, for numerical analysis, we set the renormalization scale $\Lambda$ as 300 GeV and the mass of top quark as 175 GeV.
The effective $U(1)'$ charges, ${\tilde Q}_i$, may be redefined as $Q_i = g_1^{'} {\tilde Q}_i$, for the sake of convenience, because ${\tilde Q}_i$ always go together with the $U(1)'$ gauge coupling constant [34].
Like ${\tilde Q}_i$, $Q_i$ also satisfy the $U(1)'$ gauge invariance condition, $\sum Q_i = 0$.
We have established the experimentally allowed region in the ($Q_1, Q_2$)-plane in Ref. [34], using such constraints as the extra gauge boson mass and the mixing angle between the two neutral gauge boson $\alpha_{ZZ'}$.
We adopt the results of Ref. [34], namely, $Q_1 = - 1$, $Q_2 = - 0.1$, and $Q_3 = 1.1$.
We further set for our numerical analysis $\tan\beta =3$, $s(0) = m_T = 500$ GeV, and $A_t = 100$ GeV.
Then, we are left with four free parameters: $\phi_t$, $\lambda$, $A_{\lambda}$, and $T$ in the full effective potential.

In Fig. 1, the equipotential contours of the effective potential at the one-loop level, $\langle V(v_1,v_2,T) \rangle$, are plotted in the ($v_1, v_2$)-plane for $\phi_t = \pi/1000$, $\lambda = 0.8$, $A_{\lambda} = 2271$ GeV, and $T = 100$ GeV.
We take a very small value for the phase $\phi_t$ in order to study the case of an almost exact CP conservation.
Since there are two degenerate minima in Fig. 1, $T = 100$ GeV is the critical temperature, and a first-order EWPT is possible.
Let us examine the strength of the phase transition.
The coordinates of the two minima in the ($v_1, v_2$)-plane determine the VEVs at the two degenerate vacua as: $(v_1, v_2, s) = (1,3,489)$ and $(199,507,617)$ in GeV.
Note that $s$ is obtained indirectly through the nonlinear equation.
The critical VEV is then calculated to be about $v_c = 557$ GeV for Fig. 1.
Thus, the baryon preserving condition is satisfied enough since $v_c/T_c = 5.57$.
It is therefore possible that the present model accommodates a strongly first-order EWPT.
The masses of the four neutral Higgs bosons are obtained using the parameter values in Fig. 1 as 73, 792, 1748, and 1750 GeV.

In Fig. 2, the equipotential contours of $\langle V(v_1,v_2,T) \rangle$ are plotted in the ($v_1, v_2$)-plane for $\phi_t = \pi/2$, $\lambda = 0.8$, $A_{\lambda} = 2271$ GeV, and $T = 147$ GeV.
We take the maximal value of the phase $\phi_t$ for the CP violation,
that is, $\sin\phi_t = 1$, in order to study the effect of the CP phase on the strength of the phase transition.
The critical temperature in Fig. 2 is 147 GeV, since there are two degenerate minima.
The coordinates of the minima are the same as Fig. 1, hence the critical VEV is the same as Fig. 1.
However, the critical temperature is different from Fig. 1, and therefore the strength of the phase transition is different from Fig. 1: it is $v_c/T_c = 3.79$.
We find that the strength of the phase transition is reduced.
Also, we find that the mass of the lightest neutral Higgs boson is slightly increased as $\phi_t$ is changed from $\pi/1000$ to $\pi/2$.

We continue our numerical analysis for other parameter values.
In Fig. 3, the equipotential contours of $\langle V(v_1,v_2,T) \rangle$ are plotted in the ($v_1, v_2$)-plane for $\phi_t = \pi/1000$, $\lambda = 0.7$, $A_{\lambda} = 2115$ GeV, and $T = 100$ GeV.
The strength of the first-order EWPT is estimated to be about 6.24.
The mass of the lightest neutral Higgs boson is about 79 GeV.
In Fig. 4, for $\phi_t = \pi/2$, $\lambda = 0.7$, $A_{\lambda} = 2115$ GeV, the critical temperature is about $T_c = 143$ GeV.
We find that the strength of the first-order EWPT is decreased from 6.24 to 4.36, and the mass of the lightest neutral Higgs boson is slightly increased from 79 GeV to about 80 GeV, as $\phi_t$ is changed from $\pi/1000$ to $\pi/2$, comparing the numerical results of Fig. 4 with those of Fig. 3.

The effect of $\phi_t$ that accounts for the CP mixing between the scalar and pseudoscalar Higgs bosons upon the strength of the first-order EWPT seems rather straightforward, if we compare Fig. 1 with Fig. 2, or Fig. 3 with Fig. 4.
Increasing the size of the CP mixing between the scalar and pseudoscalar Higgs bosons would result in decreasing the strength of the phase transition.
On the other hand, the relationship between the strength of the first-order EWPT and the mass of the lightest Higgs boson may be seen clearly if we compare Fig. 1 with Fig. 3 or Fig. 2 with Fig. 4, since in this way we can exclude the effect of $\phi_t$.
Comparing Fig. 1 with Fig. 3, where $\phi_t$ is fixed at $\pi/1000$, it is evident that the strength of the phase transition increases as the mass of the lightest Higgs boson increases.
This behavior is also observed if we compare Fig. 2 with Fig. 4, where $\phi_t$ is fixed at $\pi/2$.
Combining these two observations, one might expect that in the MSSM with $U(1)'$ in explicit CP violation scenario the strength of the phase transition increases if the mass of the lightest neutral Higgs boson increases or if the size of the CP mixing between the scalar and pseudoscalar Higgs bosons decreases.
We summarize in Table 1 the relevant numbers for Figs. 1-4.

\setcounter{table}{0}
\def\tablename{}{}%
\renewcommand\thetable{TABLE 1}
\begin{table}[t]
\caption{The values of relevant parameters in Figs. 1-4 which allow a strongly first-order EWPT in the MSSM with $U(1)'$ in explicit CP violation scenario, and the numerical results.
The remaining parameters are set as $\tan \beta = 3$, $s(0) = m_T = 500$ GeV, and $A_t = 100$ GeV.
The arrow in a cell indicates that the values in the cell is equal to those in the adjoining left cell.}
\begin{center}
\begin{tabular}{c|c|c|c|c} \hline\hline
 figures     & Fig. 1 & Fig. 2 & Fig. 3 &  Fig. 4  \\
\hline\hline
$\phi_t$ & $\pi/1000$   & $\pi/2$ & $\pi/1000$ & $\pi/2$   \\
\hline
$\lambda$ & 0.8  & $\leftarrow$ & 0.7 & $\leftarrow$   \\
\hline
$A_{\lambda}$ (GeV) & 2271 & $\leftarrow$   & 2115  & $\leftarrow$   \\
\hline
$T_c$ (GeV) & 100 & 147 & 100  & 143  \\
\hline
$m_{h_1}$ (GeV)  & 73  & 75  & 79  & 80      \\
\hline
$m_{h_2}$ (GeV)   & 792  & 792  & 789  & 789     \\
\hline
$m_{h_3}$ (GeV)   & 1748  & 1747  & 1579  & 1578     \\
\hline
$m_{h_4}$ (GeV)   & 1750  & 1749  & 1580  & 1579     \\
\hline
$(v_{1A}, v_{2A}, s_A)$  (GeV)  & (1,3,483) & $\leftarrow$  & (2,4,484) & $\leftarrow$   \\
\hline
$(v_{1B}, v_{2B}, s_B)$  (GeV)  & (199,507,617) & $\leftarrow$  & (230,564,639) & $\leftarrow$   \\
\hline
$v_c/T_c$ & 5.57 & 3.79 & 6.24 & 4.36     \\
\hline\hline
\end{tabular}
\end{center}
\end{table}

\section{CONCLUSIONS}

In the MSSM with $U(1)'$ in explicit CP violation scenario, at the effective one-loop level, we study the possibility of a strongly first-order electroweak phase transition.
The radiative corrections are evaluated by employing the effective potential method, taking into account the top and stop quark loops.
The explicit CP violation is triggered by a phase arising in the stop quark masses.
We have searched for the parameter space of the model to find that some region in the parameter space allows a strongly first-order electroweak phase transition.
Four sets of parameter values are selected and plotted in figures in order to show explicitly that a strongly first-order electroweak phase transition is possible indeed.

It is natural that the strength of the phase transition would change if we choose different sets of parameter values.
In particular, it may depend on the size of the phase for the CP mixing between the scalar and pseudoscalar Higgs bosons.
We find that the strength of the strongly first-order electroweak phase transition decreases as the phase for the CP mixing increases.

The mass of the lightest neutral Higgs boson is also changed as the parameter values are changed.
We find that, for a fixed value of the CP phase, the lightest neutral Higgs boson becomes heavier when the strength of the phase transition is enhanced.
In other words, if the phase transition is stronger for a set of parameter values than for another set of parameter values, then the mass of the lightest Higgs boson is also larger for the former set of parameter values than for the latter.

\vskip 0.3 in
\noindent
{\large {\bf ACKNOWLEDGMENTS}}
\vskip 0.2 in
This research is supported by KOSEF through CHEP.
The authors would like to acknowledge the support from KISTI (Korea Institute of Science
and Technology Information) under "The Strategic Supercomputing Support Program"
with Dr. Kihyeon Cho as the technical supporter.
The use of the computing system of the Supercomputing Center is also greatly appreciated.

\vfil\eject


\vfil\eject


{\large {\bf FIGURE CAPTION}}

\vskip 0.2 in
\noindent
FIG. 1. : Equipotential contours of $\langle V (v_1, v_2, T) \rangle$
in the ($v_1, v_2$)-plane for $\phi_t = \pi/1000$, $\tan \beta =3$,
$\lambda=0.8$, $A_{\lambda} = 2271$ GeV, $s(0)= m_T = 500$ GeV, and $A_t = 100$ GeV,
at the critical temperature $T = 100$ GeV.
The two minima in the ($v_1, v_2$)-plane determine the VEVs at the two degenerate vacua as:
$(v_1, v_2, s) = (1,3,489)$ and $(199,507,617)$ in GeV.
The strength of the first-order EWPT is about $v_c/T_c = 5.57$.
The masses of the four Higgs bosons are $m_{h_1} = 73$ GeV, $m_{h_2} = 792$ GeV,
$m_{h_3} = 1748$ GeV, and $m_{h_4} = 1750$ GeV.

\vskip 0.2 in
\noindent
FIG. 2. : Equipotential contours of $\langle V (v_1, v_2, T) \rangle$
in the ($v_1, v_2$)-plane for $\phi_t = \pi/2$, $\tan \beta =3$, $\lambda=0.8$,
$A_{\lambda} = 2271$ GeV, $s(0)= m_T = 500$ GeV, and $A_t = 100$ GeV, at the critical temperature $T_c = 147$ GeV.
The VEVs at the two degenerate vacua are the same as Fig. 1.
The strength of the first-order EWPT is about $v_c/T_c = 3.79$.
The masses of the four Higgs bosons are $m_{h_1} = 75$ GeV, $m_{h_2} = 792$ GeV,
$m_{h_3} = 1747$ GeV, and $m_{h_4} = 1749$ GeV.

\vskip 0.2 in
\noindent
FIG. 3. : Equipotential contours of $\langle V (v_1, v_2, T) \rangle$
in the ($v_1, v_2$)-plane for $\phi_t = \pi/1000$, $\tan \beta =3$, $\lambda=0.7$,
$A_{\lambda} = 2115$ GeV, $s(0)= m_T = 500$ GeV, and $A_t = 100$ GeV, at the critical temperature $T_c = 100$ GeV.
The two minima in the ($v_1, v_2$)-plane determine the VEVs at the two degenerate vacua as:
$(v_1, v_2, s) = (2,4,484)$ and $(230,564,639)$ in GeV.
The strength of the first-order EWPT is about $v_c/T_c = 6.24$.
The masses of the four Higgs bosons are $m_{h_1} = 79$ GeV, $m_{h_2} = 789$ GeV,
$m_{h_3} = 1579$ GeV, and $m_{h_4} = 1580$ GeV.

\vskip 0.2 in
\noindent
FIG. 4. : Equipotential contours of $\langle V (v_1, v_2, T) \rangle$
in the ($v_1, v_2$)-plane for $\phi_t = \pi/2$, $\tan \beta =3$, $\lambda=0.7$,
$A_{\lambda} = 2115$ GeV, $s(0)= m_T = 500$ GeV, and $A_t = 100$ GeV, at the critical temperature $T_c = 143$ GeV.
The VEVs at the two degenerate vacua are the same as Fig. 3.
The critical temperature is determined by $T_c = 143$ GeV.
The strength of the first-order EWPT is about $v_c/T_c = 4.36$.
The masses of the four Higgs bosons are $m_{h_1} = 80$ GeV, $m_{h_2} = 789$ GeV,
$m_{h_3} = 1578$ GeV, and $m_{h_4} = 1579$ GeV.

\vfil\eject

\setcounter{figure}{0}
\def\figurename{}{}%
\renewcommand\thefigure{FIG. 1}
\begin{figure}[t]
\begin{center}
\includegraphics[scale=0.6]{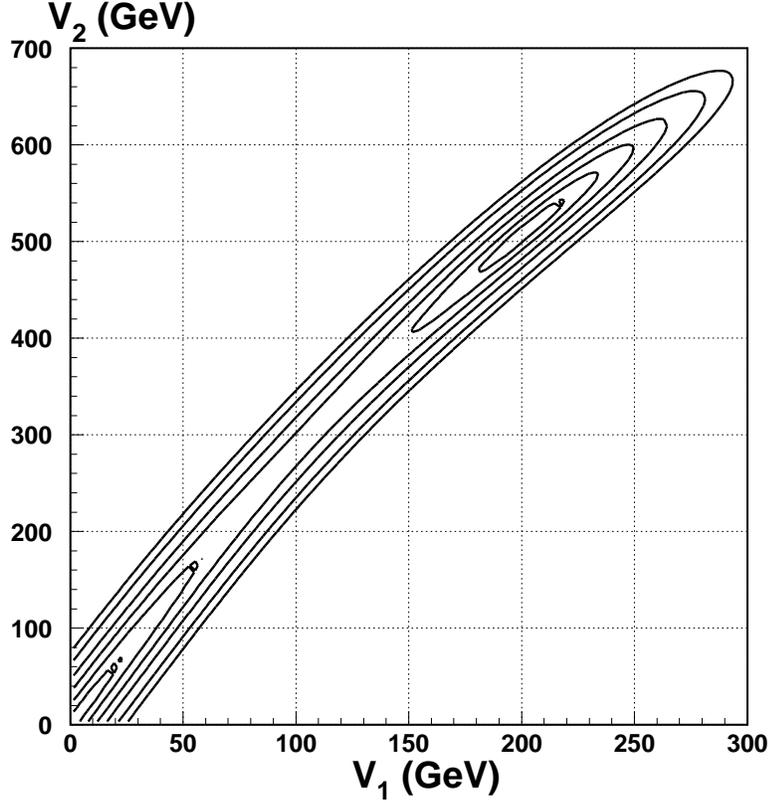}
\caption[plot]{Equipotential contours of $\langle V (v_1, v_2, T) \rangle$
in the ($v_1, v_2$)-plane for $\phi_t = \pi/1000$, $\tan \beta =3$, $\lambda=0.8$,
$A_{\lambda} = 2271$ GeV, $s(0)= m_T = 500$ GeV, and $A_t = 100$ GeV, at the critical temperature $T = 100$ GeV.
The two minima in the ($v_1, v_2$)-plane determine the VEVs at the two degenerate vacua as:
$(v_1, v_2, s) = (1,3,489)$ and $(199,507,617)$ in GeV.
The strength of the first-order EWPT is about $v_c/T_c = 5.57$.
The masses of the four Higgs bosons are $m_{h_1} = 73$ GeV, $m_{h_2} = 792$ GeV,
$m_{h_3} = 1748$ GeV, and $m_{h_4} = 1750$ GeV.}
\end{center}
\end{figure}

\setcounter{figure}{0}
\def\figurename{}{}%
\renewcommand\thefigure{FIG. 2}
\begin{figure}[t]
\begin{center}
\includegraphics[scale=0.6]{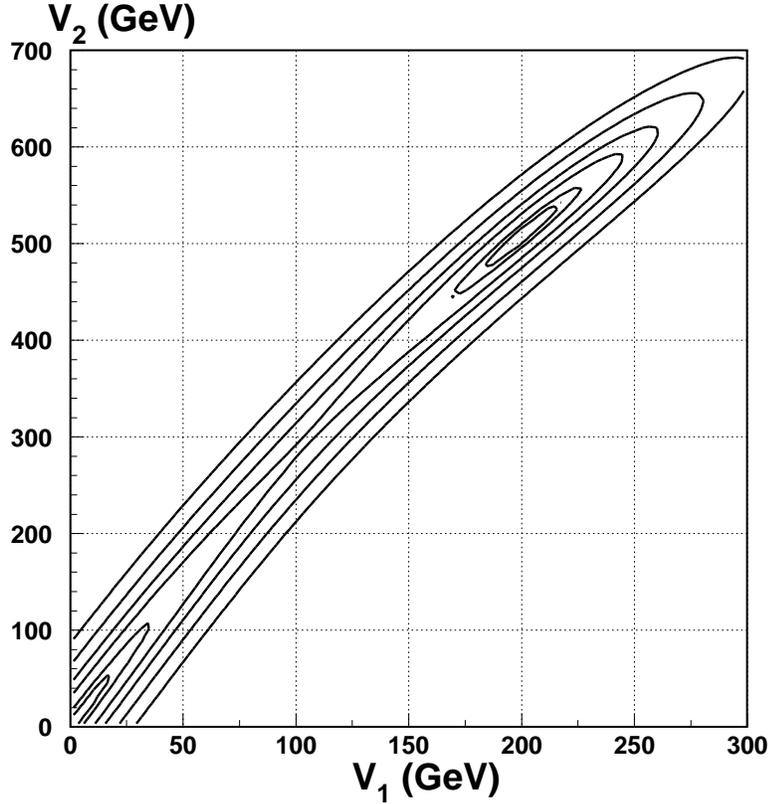}
\caption[plot]{Equipotential contours of $\langle V (v_1, v_2, T) \rangle$
in the ($v_1, v_2$)-plane for $\phi_t = \pi/2$, $\tan \beta =3$, $\lambda=0.8$,
$A_{\lambda} = 2271$ GeV, $s(0)= m_T = 500$ GeV, and $A_t = 100$ GeV, at the critical temperature $T_c = 147$ GeV.
The VEVs at the two degenerate vacua are the same as Fig. 1.
The strength of the first-order EWPT is about $v_c/T_c = 3.79$.
The masses of the four Higgs bosons are $m_{h_1} = 75$ GeV, $m_{h_2} = 792$ GeV,
$m_{h_3} = 1747$ GeV, and $m_{h_4} = 1749$ GeV.}
\end{center}
\end{figure}

\setcounter{figure}{0}
\def\figurename{}{}%
\renewcommand\thefigure{FIG. 3}
\begin{figure}[t]
\begin{center}
\includegraphics[scale=0.6]{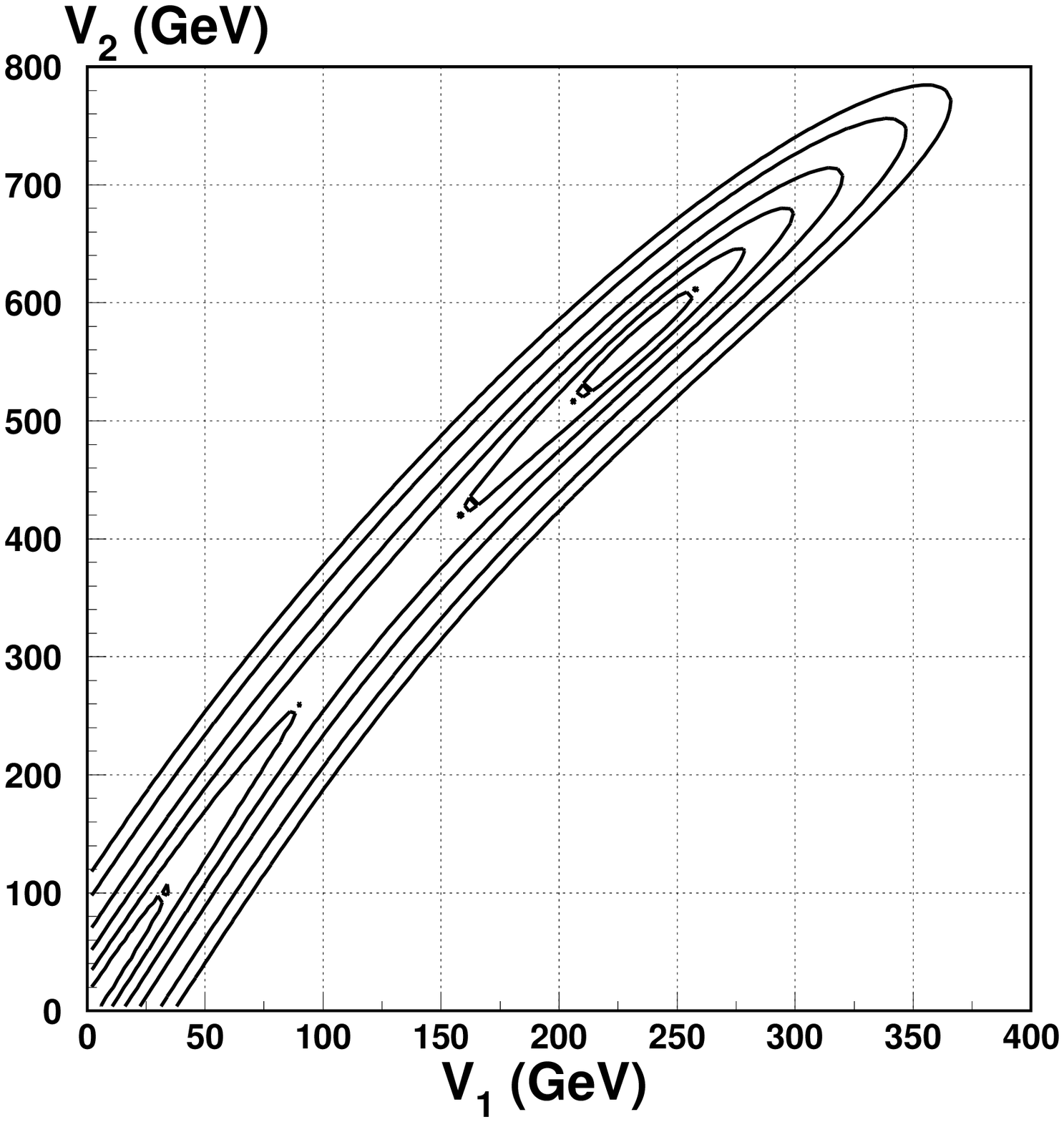}
\caption[plot]{Equipotential contours of $\langle V (v_1, v_2, T) \rangle$ in the ($v_1, v_2$)-plane
for $\phi_t = \pi/1000$, $\tan \beta =3$, $\lambda=0.7$, $A_{\lambda} = 2115$ GeV,
$s(0)= m_T = 500$ GeV, and $A_t = 100$ GeV, at the critical temperature $T_c = 100$ GeV.
The two minima in the ($v_1, v_2$)-plane determine the VEVs at the two degenerate vacua as:
$(v_1, v_2, s) = (2,4,484)$ and $(230,564,639)$ in GeV.
The strength of the first-order EWPT is about $v_c/T_c = 6.24$.
The masses of the four Higgs bosons are $m_{h_1} = 79$ GeV, $m_{h_2} = 789$ GeV,
$m_{h_3} = 1579$ GeV, and $m_{h_4} = 1580$ GeV.}
\end{center}
\end{figure}

\setcounter{figure}{0}
\def\figurename{}{}%
\renewcommand\thefigure{FIG. 4}
\begin{figure}[t]
\begin{center}
\includegraphics[scale=0.6]{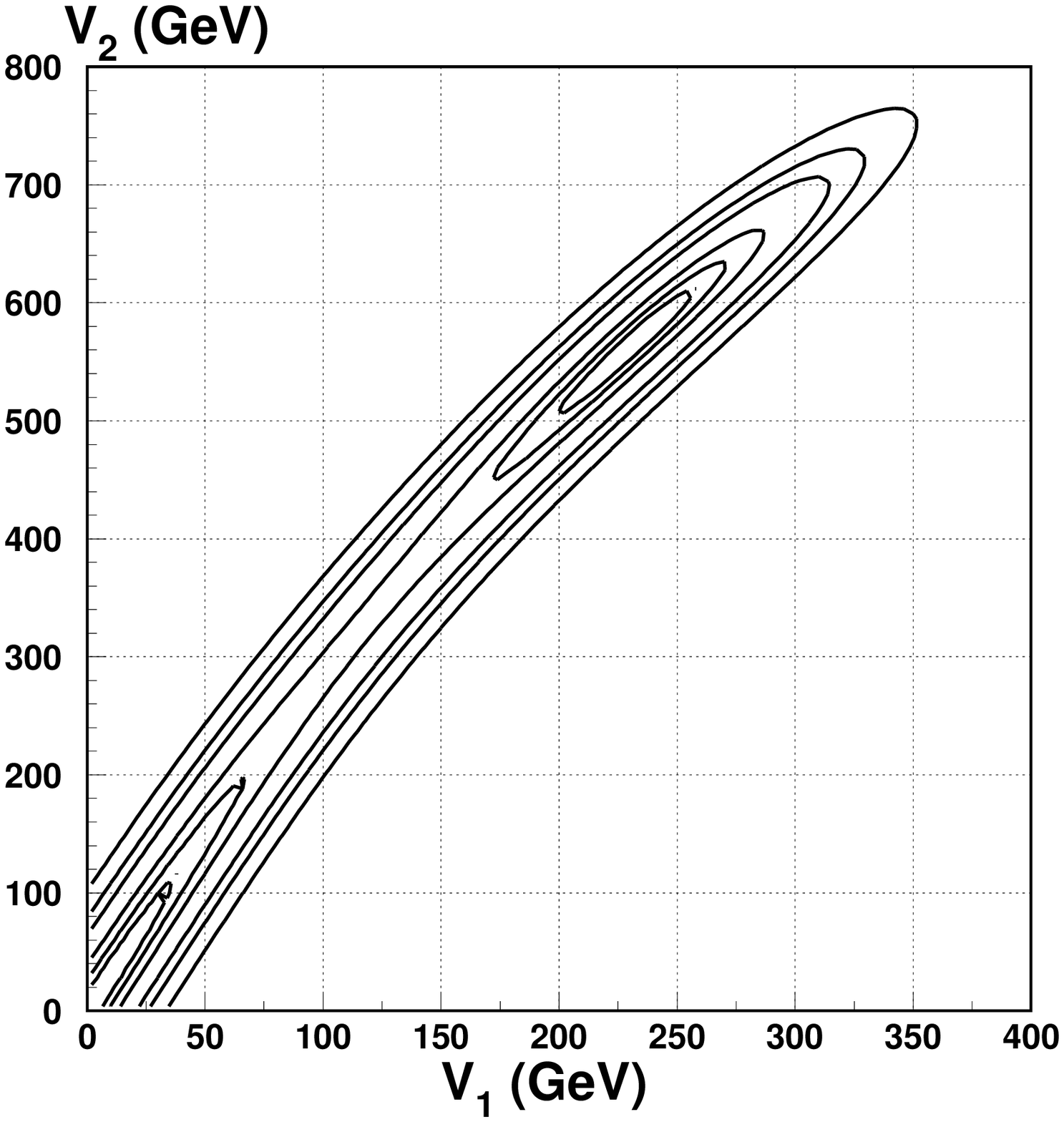}
\caption[plot]{Equipotential contours of $\langle V (v_1, v_2, T) \rangle$
in the ($v_1, v_2$)-plane for $\phi_t = \pi/2$, $\tan \beta =3$, $\lambda=0.7$,
$A_{\lambda} = 2115$ GeV, $s(0)= m_T = 500$ GeV, and $A_t = 100$ GeV,
at the critical temperature $T_c = 143$ GeV.
The VEVs at the two degenerate vacua are the same as Fig. 3.
The critical temperature is determined by $T_c = 143$ GeV.
The strength of the first-order EWPT is about $v_c/T_c = 4.36$.
The masses of the four Higgs bosons are $m_{h_1} = 80$ GeV, $m_{h_2} = 789$ GeV,
$m_{h_3} = 1578$ GeV, and $m_{h_4} = 1579$ GeV.}
\end{center}
\end{figure}


\begin{thebibliography}{99}
\bibitem{1} A. D. Sakharov, JETP Lett. {\bf 5}, 24 (1967).
\bibitem{2}  M. E. Shaposhnikov, JETP Lett. {\bf 44}, 465 (1986); Nucl. Phys. B {\bf 287}, 757 (1987);
 Nucl. Phys. B {\bf 299}, 797 (1988).
\bibitem{3} S. Barr, G. Segre, and H. A. Weldon, Phys. Rev. D {\bf 20}, 2494 (1979).
\bibitem{4} G. R. Farrar and M. E. Shaposhnikov, Phys. Rev. D {\bf 50}, 774 (1994).
\bibitem{5} S. Weinberg, Phys. Rev. Lett. {\bf 37}, 657 (1976).
\bibitem{6} J. R. Espinosa, M. Quir\'os, and F. Zwirner, Phys. Lett. B {\bf 307}, 106 (1993).
\bibitem{7} M. Carena, M. Quir\'os and C. E. M. Wagner, Phys. Lett. B {\bf 380}, 81 (1996); Nucl. Phys. B {\bf 524}, 3 (1998).
\bibitem{8} D. Delepine, J. M. Gerard, R. G. Felipe, and J. Weyers, Phys. Lett. B {\bf 386}, 183 (1996).
\bibitem{9} M. Laine and K. Rummukainen, Phys. Rev. Lett. {\bf 80}, 5259 (1998);
Nucl. Phys. B {\bf 535}, 423 (1998).
\bibitem{10} J. M. Cline and G. D. Moore, Phys. Rev. Lett.  {\bf 81}, 3315 (1998).
\bibitem{11} K. Funakubo, A. Kakuto, S. Otsuki, and F. Toyoda, Prog. Theor. Phys. {\bf 99}, 1045 (1998);
K. Funakubo, Prog. Theor. Phys. {\bf 101}, 415 (1999).
\bibitem{12} M. Losada, Nucl. Phys. B {\bf 537}, 3 (1999).
\bibitem{13} F. Csikor, Z. Fodor, P. Hegedus, A. Jakovac, S. D. Katz, A. Piroth,
Phys. Rev. Lett. {\bf 85}, 932 (2000).
\bibitem{14} S. J. Huber, T. Konstandin, T. Prokopec, and M. G. Schmidt, Nucl. Phys. A {\bf 785}, 206 (2007).
\bibitem{15} M. Pietroni, Nucl. Phys. B {\bf 402}, 27 (1993).
\bibitem{16} M. Bastero-Gil, C. Hugonie, S .F. King, D. P. Roy, and S. Vempati, Phys. Lett. B {\bf 489}, 359 (2000).
\bibitem{17} A. T. Davies, C. D. Froggatt, R. G. Moorhouse, Phys. Lett. B {\bf 372}, 88 (1996).
\bibitem{18} S. J. Huber and M. G. Schmidt, Eur. Phys. J. C {\bf 10}, 473 (1999); Nucl. Phys. B {\bf 606}, 183 (2001).
\bibitem{19} A. Menon, D.E. Morrissey, and C.E.M. Wagner, Phys. Rev. D {\bf 70}, 035005 (2004).
\bibitem{20} J. Kang, P. Langacker, T. Li, and T. Liu, Phys. Rev. Lett. {\bf 94}, 061801 (2005).
\bibitem{21} S.J. Huber, T. Konstandin, T. Prokopec, and M.G. Schmidt, Nucl. Phys. B {\bf 757}, 172 (2006).
\bibitem{22} S. W. Ham, S. K. Oh, C. M. Kim, E. J. Yoo, and D. Son, Phys. Rev. D {\bf 70}, 075001 (2004);
             S.W. Ham, J.O. Im, and S.K. OH, arXiv:hep-ph/0707.4543.
\bibitem{23} S. W. Ham, S. K. Oh, and D. Son, Phys. Rev. D {\bf 71}, 015001 (2005).
\bibitem{24} S. W. Ham, E. J. Yoo, and S. K. OH, arXiv:hep-ph/0704.0328.
\bibitem{25} J. L. Hewett and T. G. Rizzo, Phys. Rep. {\bf 183}, 193 (1989).
\bibitem{26} A. Leike, Phys. Rep. {\bf 317}, 143 (1999).
\bibitem{27} M. Cvetic and P. Langacker, Phys. Rev. D {\bf 54}, 3570 (1996).
\bibitem{28} M. Cvetic, D. A. Demir, J. R. Espinosa, L. Everett, and P. Langacker, Phys. Rev. D {\bf 54}, 3570 (1996); Phys. Rev. D {\bf 56}, 2861 (1997); Erratum-ibid. D {\bf 58}, 119905 (1998).
\bibitem{29} D. A. Demir and N. K. Pak, Phys. Rev. D {\bf 57}, 6609 (1998).
\bibitem{30} Y. Daikoku and D. Suematsu, Phys. Rev. D {\bf 62}, 095006 (1998).
\bibitem{31} H. Amini, New J. Phys. {\bf 5}, 49 (2003).
\bibitem{32} S. F. King, S. Moretti, and R. Nevzorov, Phys. Rev. D {\bf 73}, 035009 (2006); Phys. Lett. B {\bf 634}, 278 (2006).
\bibitem{33} D. A. Demir and L. L. Everett, Phys. Rev. D {\bf 69}, 015008 (2004).
\bibitem{34} S. W. Ham, E. J. Yoo, and S. K. Oh, hep-ph/0703041.
\bibitem{35} S. Coleman and E. Weinberg, Phys. Rev. D {\bf 7}, 1888 (1973).
\bibitem{36} L. Dolan and R. Jackiw, Phys. Rev. D {\bf 9}, 3320 (1974).
\end{thebibliography}
\end{document}